# A Privacy Preserving IoT Data Marketplace Using IOTA Smart Contracts

Hadi Farahani, Hamid Reza Shahriari
Atlas Lab, Department of Computer Engineering
Amirkabir University of Technology (Tehran Polytechnic)
Tehran, Iran
Email: {hadifarahani, shahriari}@aut.ac.ir

## Abstract

In recent years, the volume of data generated by IoT devices has increased dramatically. Using this data can improve decision-making in the public and private sectors and increase productivity. Many attempts have been made to enhance and adapt businesses to exploit this IoT data. Among these, IoT data trading is the most popular approach. To this end, ongoing projects are currently focused on developing decentralized data marketplaces for IoT using blockchain and cryptocurrencies. Here we explore how a decentralized data marketplace could be created using IOTA tangle and IOTA smart contract chains (SC chains). We also consider the advantages of such architecture in terms of cost, scalability, and privacy over current designs and introduce the various elements it should have.

## I. INTRODUCTION

Internet of Things (IoT) envisions connecting billions of devices and sensors to the Internet. In recent years, the number of devices in IoT has grown at a steady pace. As a result, the amount of data generated by these devices has grown exponentially. A new forecast from IDC estimates that there will be 55.9 billion connected IoT devices, generating 79.4 ZB of data in 2025. Using this data can lead to efficient and effective resource management. For example, in Smart Cities, we have critical issues like Waste, traffic, energy, water, education, unemployment, health, and crime management. Using this data can lead to inform decision-making and reduce costs, eliminate Waste, and increase productivity [1]-[3].

One practical way to use this data would be to exchange it in a decentralized data marketplace, open to everyone, and use cryptocurrencies to pay for data. We need a secure, cheap, fast, and scalable platform to facilitate data trades in such a marketplace without violating users' privacy. To this end, several researches have explored such marketplaces[3]-[9].

The authors in [3,4,5,6] presented their blockchain-based platforms for data trading and proposed using the Ethereum blockchain and its smart contracts to develop this marketplace. However, these platforms do not operate cost-effectively, as the cost of system operations in a trade exceeds the exchanged value. In general, using Ethereum as a blockchain platform is not economically justifiable due to its rising transaction fees. Additionally, users' privacy is not considered in the design of these platforms. Another issue is that the throughput of their platforms is limited by the low scalability of the Ethereum network.

The authors In [7] presented a system architecture for data sharing in smart transportation systems, in which IOTA tangle, IPFS, and Ethereum smart contracts are used for data storage and coordinating the data sharing. But similarly, their design is not cost-effective and scalable due to the use of Ethereum.

In [8], the authors proposed using the Lightning network to fulfill the requirement of having micro-payment and cost-effective system operation. However, establishing and closing a payment channel in the lightning network requires two transactions, namely funding transaction, and closing transaction in the Bitcoin network. Due to the high transaction fee in the Bitcoin blockchain, this proposal is not cost-effective. Even if the seller and buyer use their already-established channels, their privacy would be in danger. Generally, privacy is in contrast with efficiency (efficient routing) in the lightning network.

In [9], the authors proposed using Ethermint (instead of the Ethereum network) and state channels to develop the data marketplace. But they did not consider users' privacy in their design.

In an IoT data marketplace, we have a large volume of trades, so the scalability of this market should be considered in the designed platforms. Also, the payments in a data marketplace, trading sensor data, are micro-payments. Hence, using public blockchains like Bitcoin and Ethereum, due to their low scalability (transaction throughput) and high transaction fees, is not suitable for this purpose.

In this paper, we have chosen the IOTA tangle due to its feeless and scalable nature, as well as IOTA smart contracts to implement the business logic of the marketplace in a privacy-preserving manner. We also used a decentralized data storage platform to store and access sensor data and a decentralized certificate Authority scheme to ensure data authenticity.

The rest of this paper is organized as follows: Section II presents the general architecture of our system. The implementation and proposed protocol are presented in section III. The performance analysis of our platform is discussed in section IV. Section V concludes this paper.

## II. SYSTEM ARCHITECTURE

To address the shortcomings of existing data marketplaces and design a platform that meets our requirements, multiple distributed technologies are used, Such as IOTA, decentralized data storage (IPFS), and distributed certificate authority (see figure 1). The following introduces the entities, adversary model, requirements, and the role of each technology in the data marketplace.

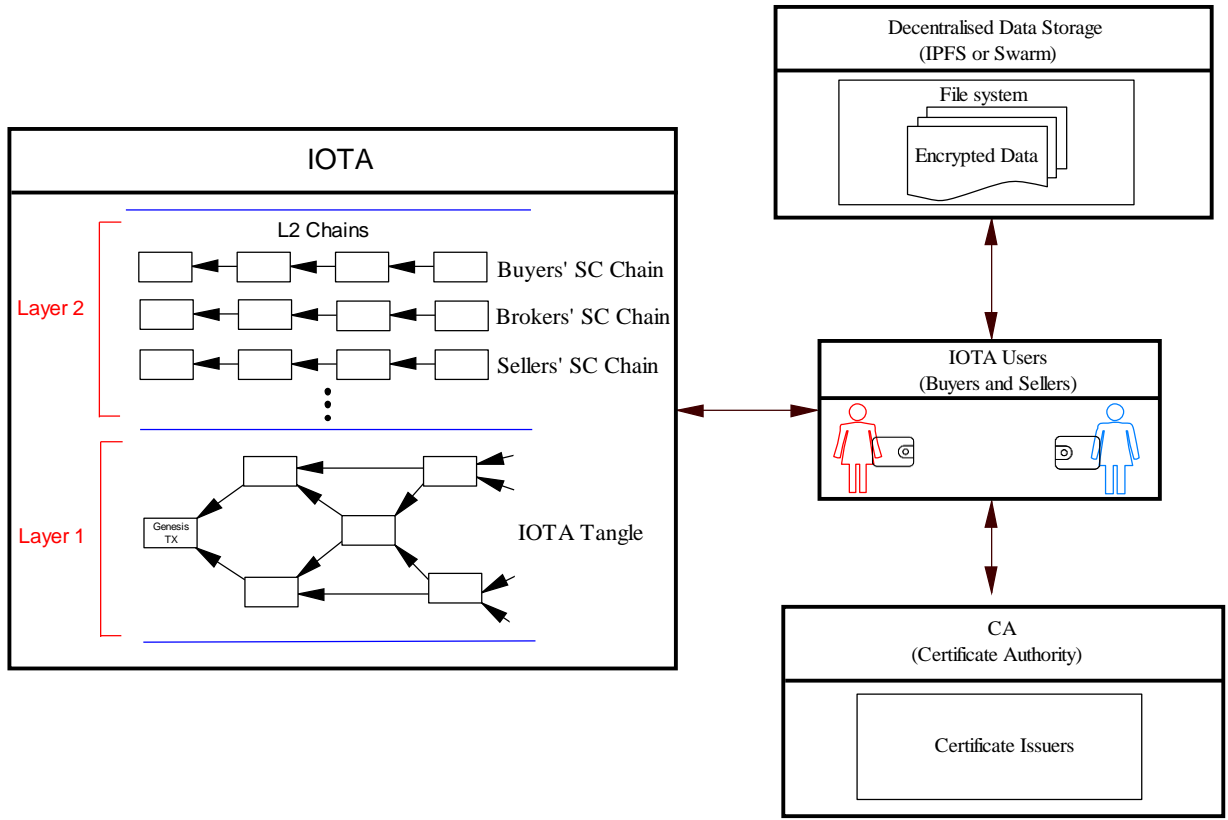

Fig. 1. Infrastructure Architecture Diagram

### A. Platform Entities

Three types of entities take part in the platform.
1) Sellers: who wish to sell their sensor data and profit from it. They use their wallet to transact with the marketplace.
2) Buyers: who wish to buy their required data(a specific type of sensor data) and use it to increase productivity and efficiency in their own particular application. They use their wallet to transact with the marketplace.
3) Brokers: trusted intermediaries who match sellers and buyers and maintain the functions of the data marketplace, which are represented as smart contracts deployed on the brokers' SC chain and thus are uniquely identified by their SC address.

Buyers and sellers are users of the IOTA tangle and can act as a validator of any chain, including Sellers' and buyers' SC chains. However, they cannot be a validator of the brokers' SC chain. The brokers' SC chain is an exceptional chain in which the chain validators are considered trusted. These validators have their real-world reputations, and collusion among them is highly unlikely due to fierce competition between them [11]. For example, regulatory agencies can be a part of these validators.

### B. Adversary model

All entities are not trusted except the validators of the broker's chain and certificate authority(CA). The attackers can be either internal or external and may attempt to jeopardize the privacy of users and gain information about them and their trades. The internal attacker (who can be either a buyer or a seller) aims to find out the identity of the corresponding party in his or her data deals. The external attacker is a third party who wishes to gain information about marketplace transactions and violate the users' privacy. Since the attacker can be a buyer or seller, he or she can be a validator of the Sellers or buyers' SC chain and thus has access to the block contents of mentioned chains. Also, they can query validators of the brokers' SC chain or peruse the anchoring transactions [11] and access the block contents of this chain. Still, they cannot be a validator of the brokers' chain and interfere in the brokers' SC chain operations.

C. Platform and User Requirements

1) User privacy: It is required that users (sellers or buyers) or any third party cannot infer who bought what data from whom(which buyer purchased what data from which seller). In other words, matching sellers and buyers in data trades and tracing communications must be inaccessible to a third party. Further, no adversary can find out the identity and the exact location of sellers and buyers.
2) Data Authenticity: it should be possible for the platform (brokers) to check the authenticity of data and the seller who collected the data.
3) Cost-effective performance (Transactional costs): since payments in the marketplace are micropayments, the cost of the system's operation must be significantly low to make the use of a decentralized marketplace economically justifiable.
4) System scalability (High Transaction throughput): The system should have acceptable speed and respond in a reasonable time so it can handle a lot of requests (trades). Public blockchains like Bitcoin have a very low transaction throughput and thus are not suitable for our purpose.

D. The Function and Role of Each Technology in The Data Marketplace

1) IOTA:

We use IOTA as a platform that enables interactions between different entities in the marketplace and enforces the business logic of the data marketplace using its smart contracts [10], [11].

On layer one (L1), we have the tangle, which is a DAG-based ledger. It acts as a trust and security anchor and interoperability framework for layer two (L2) SC chains.

On L2, we have multiple smart contract chains. Each chain is a separate blockchain with smart contracts on it, functionally equivalent to Ethereum smart contracts and fully composable between each other. Using this multi-chain environment enables IOTA to achieve scalability due to massive parallelization through multiple parallel ledgers (also known as shards) and parallel smart contracts [11]. We assume that these chains are used for many purposes, and the data marketplace is one of them.

In our system, we used this multi-chain environment, but in order to explain the protocol and system flow simply, we only used three L2 chains and named them sellers' SC chain, Buyers' SC chain, and Brokers' SC chain. The first and the second are considered public blockchains and permissionless, but the latter is considered a consortium blockchain and permissioned.

We use these specified chains to explain how the interactions in this marketplace could take place in a privacy-preserving manner. The function of the Brokers' SC chain is to separate the events (relations between transactions) of sellers' SC chain from buyers' SC chain and, indeed, prevent adversaries from any inference about the market's exchanges.

We hypothesize that the seller and buyer in a data deal are on different chains. However, sellers and buyers in this market might be on any chain on L2 except the smart contract chain used by the broker. Even if the buyer and seller were on the same chain, their privacy would be guaranteed due to indirect communication through the broker's SC.

It is also worth mentioning that we can have multiple smart contracts on the Brokers' SC chain, each responsible for a specific type of data, and it is not just one contract.

2) Certificate Authority:

In this platform, we used an infrastructure analogous to PKI to ensure data authenticity. We call this infrastructure CA, in which there are a large number of certificate issuers (CI), organized in a distributed and decentralized manner, to verify the seller's data authenticity and issue a certificate for her.

A Seller communicates with a CI in a privacy-preserving manner and declares its approximate location and the type of sensors she possesses. She also sends a small amount of her sensor data and a new nonce to the CA. Then, the seller will receive a signed certificate from the CA, which certifies that the holder of this certificate is present in a specific location and has a particular type of sensor. This certificate has a timestamp and expiration date and time as well as the new nonce generated by the seller in order to prevent replay attacks.

Certificate issuers can use AI and machine learning techniques to verify the authenticity of data generated by each type of sensor in any geographical location.

In order to implement this infrastructure, we can use the solutions proposed by researchers in the literature. There are many researches for determining the geolocation of IoT devices without violating privacy. For example, authors In [6] proposed using GeoHex to protect individuals' location privacy. In [7], the authors proposed using FOAM, an open protocol for decentralized and geospatial data markets, in conjunction with Zero Knowledge proof of location. Since many articles have addressed this issue, We suffice to mention the function of the CA and not go into the details about CA's implementation.

3) Decentralised Data Storage technology:

Here, we used decentralized data storage platforms like IPFS or Swarm to store and access sensor data. This storage acts as an intermediary between buyers and sellers to prevent direct communication and leaking IP addresses. In this marketplace, the seller would encrypt the data and store it in this storage, and then the buyer can retrieve the encrypted data from storage, Instead of the seller sending the data directly to the buyer.

## III. IMPLEMENTATION

This section presents the designed protocol to illustrate the interactions between different entities in the Data marketplace. We assumed that (i) the address of the broker's smart contract is publicly known; (ii) the seller contacted CA beforehand and obtained the certificate; and (iii) the smart contracts have been deployed beforehand by their associated entities. Table 1 contains the notations used in this protocol, and figure 2 shows how entities interact during a trade:

Table 1. Parameters used in protocol.

| Parameter | Description |
| --- | --- |
| DD | Detailed Description of collected/required Data |
| $KU_{Buyer}$ | Buyer's Public key (different in each trade) |
| $KU_{Seller}$ | Seller's Public key (different in each trade) |
| $K_s$ | Symmetric Key used to encrypt/decrypt sensor data |
| $E_{KU_{Buyer}}$ | Encryption with Buyer's public key |
| $E_{KU_{Seller}}$ | Encryption with Seller's public key |
| $ID_s$ | Fresh nonce generated by seller |
| $ID_b$ | Fresh nonce generated by buyer |

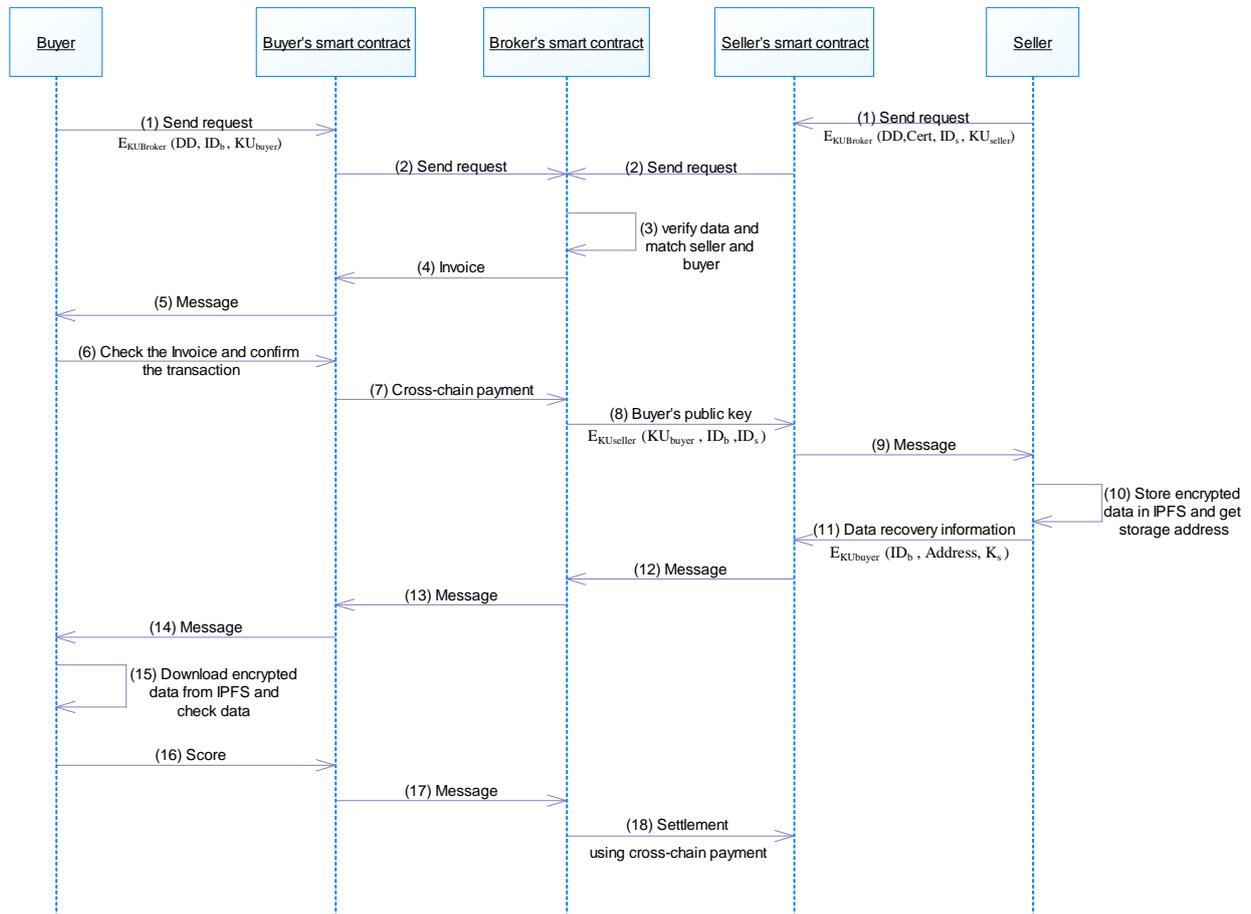

Fig. 2. Outline of Our Protocol (sequence of transactions)

1) First, The seller generates a pair of public/private keys for this trade. Second, she generates a request consisting of collected data description, a fresh nonce, certificate, and her public key, all encrypted in the broker's public key. She sends this request to the smart contract on the seller's SC chain that she deployed beforehand, using her wallet. Also, The buyer generates his public/private key pair for this trade and then generates a request consisting of the required data description, a fresh nonce, and her public key, all encrypted in the broker's public key. She sends this request to the smart contract on the buyer's SC chain that she deployed beforehand, using her wallet:

$$E_{KU_{Broker}}(DD, Cert, ID_s, KU_{seller})$$
$$E_{KU_{Broker}}(DD, ID_b, KU_{buyer})$$

2) The smart contracts on the sellers' SC chain and buyers' SC chain wrap these requests into the UTXO and send them to the broker's smart contract on the brokers' SC chain through L1.
3) The validators of the brokers' SC chain process these requests by executing the contract. The contract decrypts these requests with its private key, stored in the secure enclave of each validator node. Then, the contract (1) verifies the data authenticity by checking the seller's certificate; (2) matches the buyer and seller based on the descriptions of their required and collected data; and (3) calculates the price of the data. Here we assume that there are different types of sensors; which basic price of each type is known to all users, and the price of every data item can be calculated by the following formula:

Price = basic price of sensor * volume of data

This step can be modified by allowing the seller to choose between different buyers based on their use cases.

4) The contract issues the invoice, encrypts it using the buyer's public key, and then sends it to the buyer's contract on the buyer's SC chain through L1.
5) The smart contract on the buyer's SC chain passes the encrypted message containing the invoice to the buyer.
6) The buyer decrypts the message, checks to see if everything is okay, and then confirms the transaction.
7) The buyer's smart contract pays the invoice using cross-chain payment (L1 asset transfers between chains).
8) The broker's smart contract generates a message consisting of the buyer's public key, the buyer's nonce, and the seller's nonce, all encrypted in the seller's public key, and sends this message to the seller's smart contract on the seller's SC chain:

$$E_{KU_{seller}}(KU_{buyer}, ID_b, ID_s)$$

9) The seller's smart contract sends this encrypted message to the seller.
10) First, The seller decrypts the message using her private key, checks to see if her nonce is in the message, and obtains the buyer's public key. Second, she encrypts her sensor data using a symmetric key (Ks) and stores the encrypted data in IPFS (or Swarm):

$$E_{ks}(Data)$$

11) After The seller stored the data and got the storage address, she generates a message consisting of the buyer's nonce, the address of encrypted data in IPFS, and the symmetric key, all encrypted in the buyer's public key. Then, she sends this message to the smart contract on the seller's SC chain:

$$E_{KU_{buyer}}(ID_b, Address, K_s)$$

12) The seller's smart contract wraps this encrypted message into the UTXO and sends it to the broker's smart contract on the brokers' SC chain through L1.
13) Similarly, the broker's smart contract conveys this message to the buyer's smart contract on the buyers' SC chain.
14) The buyer's smart contract sends this message to the buyer.
15) The buyer decrypts this message using her private key and obtains the symmetric key (Ks) and the address of encrypted data in IPFS. Then, she downloads the encrypted data from IPFS and decrypts it using Ks.
16) The buyer calculates the seller's score given the data she received, encrypts, and sends this score to her smart contract on the buyer's SC chain.
17) The buyer's smart contract sends this encrypted score to the broker's smart contract.
18) The broker's smart contract updates its table, and then settles with the seller using cross-chain payment (L1 asset transfers between chains).

We assume that the broker uses a reputation system and stores all the scores of sellers in a table. Each row of this table contains a smart contract ID and the associated score.

Note that the broker can settle with the seller after receiving the message in step 12 and does not have to wait if the buyer fails to follow the protocol by doing steps 16 and 17.

In this protocol, all parties must first encrypt their messages and then send them in the form of transactions. We use freshly-generated nonces ($ID_s$, $ID_b$) to prevent Replay and modification attacks. Also, we can use padding in order to prevent adversaries from inferring message contents from the message size.

## IV. PERFORMANCE ANALYSIS

### A. Cost

Since we use L1 as a security and trust anchor and due to the anchoring mechanism of IOTA, we do not need a costly and time-consuming consensus mechanism on L2 chains like POW in the Bitcoin blockchain. Consequently, there is no need for high transaction fees and block rewards for L2 validators. The cost of operations on L2 chains is reasonably low.

In this ecosystem, users (sellers and buyers) use off-ledger requests to communicate with smart contracts on L2 SC chains. The off-ledger requests are sent by users from their wallets directly to the access nodes of the chain. It is an API call, not a transaction on L1. On the other hand, the communications between smart contracts on different L2 chains are through L1 and on-chain requests. An on-ledger request is a call to the smart contract wrapped into the UTXO. It means the on-ledger request is a transaction on L1 [11].

The off-ledger requesting is much faster and much less costly than requesting on-ledger. The off-ledger requesting makes high throughput (many hundreds of TPS on one chain) a norm[11]. Consequently, the communications between users and smart contracts are cheap and fast. Also, the L1 ledger is scalable and feeless, so the cost of communications between smart contracts on different chains is reasonably low.

### B. Scalability

Decentralized Technologies like Swarm and IPFS have acceptable latency, and they are not the bottleneck of speed and scalability in our system. So the scalability and throughput of our system depend on the functions of IOTA on L1 and L2.

Due to the relatively small number of validators on the brokers' SC chain and its consortium nature, we expect the general properties of consortium blockchains, such as increased transaction speed, scalability, and privacy, to be maintained here as well. In addition, we know the L1 ledger (IOTA tangle) is very scalable. Hence, we can have an idea about the scalability of the presented system as a whole.

### C. Privacy

As the IOTA SC team mentioned in [11], the blocks in L2 chains do not contain block headers nor transactions, only mutations of key/value pairs. This means the attacker cannot map a seller and a buyer in a data deal even if she has access to these key/value pairs. Without the transactions of the brokers' SC chain, the attacker cannot relate the events (transactions) of the buyers' SC chain to the events (transactions) of the sellers' SC chain. Further, since all requests are encrypted, the attacker cannot deduce anything about the exchanges in the marketplace, and users can participate without worrying about their privacy.

## V. DISCUSSION AND CONCLUSION

There is a trade-off between privacy and scalability (throughput). Compare the presented system with a marketplace that has only one specific chain on L2, and all sellers and buyers use this chain and its smart contracts to trade with each other. Clearly, this marketplace has higher throughput, but tracing the communications would be much easier for the attacker. Hence, the privacy of users would be jeopardized, and the attacker can gain information about users and their trades using methods similar to de-anonymization attacks.

In general, sensor data is not sensitive. But they can be sensitive in some cases, where the attacker uses the sensor data belonging to a specific user to infer sensitive information about that person and use this information to launch an attack against them. So it is important that we protect the user's privacy (identity and location) to prevent such attacks.

As mentioned in [11], the IOTA DLT is feeless at its core, as are the smart contracts on L2 chains. However, for quasi-Turing complete computations, we need the concept of gas to restrict abuse of the processor and storage resources, to prevent runaway programs and the generating of garbage in the state. To address this problem, We can fix a small amount of gas price (compared to the price of sensor data) or set a limit on the number of transactions a user can send (rate limitation strategies). At the time of this research, we don't have an exact idea of how to reward validators and how much gas and fees will be charged. We will address this problem in our future works.

In this paper, we have presented the architecture of a decentralized data marketplace that could satisfy our requirements, such as privacy, scalability, and cost-effective performance. To this end, we used technologies like IOTA, IPFS, and a Decentralized certificate authority. We demonstrated how L2 IOTA smart contract chains could be exploited to achieve our privacy goals and scalability. We have also proposed a secure and efficient protocol to illustrate how interactions in the such marketplace could take place.